# Switching Anomalous Hall and Nernst Responses by Nonmagnetic N Occupation at Fixed Noncoplanar Mn Antiferromagnetic Order

Xin Liu[1], Jiyuan Xu[1], Li Ma[1,*], Guoke Li[2,] , Dewei Zhao[2], Congmian Zhen[1], and Denglu Hou[1]

[1]*Hebei Key Laboratory of Photophysics Research and Application, College of Physics, Hebei Normal University, Shijiazhuang, 050024, China.*

[2]*Hebei Advanced Thin Films Laboratory, College of Physics, Hebei Normal University, Shijiazhuang, 050024, China.*

**Abstract:**

Nonmagnetic atomic occupation can control anomalous transverse transport by modifying magnetic symmetry without changing the underlying magnetic order. We demonstrate this effect using controlled $\gamma$-Mn, $Mn_4N$, and MnN reference states with the same lattice constant and identical noncoplanar all-in-all-out Mn magnetic configurations, while varying only the occupation of the N sublattice. The anomalous Hall and anomalous Nernst responses exhibit a pronounced zero–finite–zero evolution across the series despite the unchanged Mn spin order. In $\gamma$-Mn and MnN, the high magnetic symmetry enforces complete cancellation of the Brillouin-zone-integrated Berry curvature. In $Mn_4N$, N occupation lowers the magnetic symmetry while preserving inversion and breaking the relevant twofold rotational symmetries, thereby lifting the cancellation constraint and permitting an uncompensated Berry-curvature contribution along the [111] direction. The resulting finite anomalous Hall conductivity reaches -126 S/cm near the Fermi level. These results establish nonmagnetic sublattice occupation as a symmetry-control parameter for Berry-curvature-driven transport in compensated antiferromagnets, independent of changes in the magnetic order.

* Corresponding author. E-mail address: majimei@126.com.

## Introduction

Compensated antiferromagnets have emerged as an important platform for spintronics because they combine vanishing net magnetization and negligible stray fields with ultrafast magnetic dynamics and robustness against external magnetic perturbations[1-4]. Recent symmetry-based classifications have further broadened this landscape to unconventional antiferromagnets that exhibit ferromagnet-like electronic and transport responses despite magnetic compensation[5,6]. In particular, anomalous-Hall antiferromagnets provide electrically accessible time-reversal-odd responses without requiring a macroscopic magnetization[5-8]. Related anomalous Nernst responses have also been observed in chiral antiferromagnets, demonstrating that both electrical and thermoelectric transverse transport can survive in nearly magnetization-free states[9]. Noncollinear and noncoplanar compensated antiferromagnets therefore provide a natural setting for unconventional transverse transport[7-11].

The presence of a nontrivial magnetic order, however, does not by itself guarantee a finite macroscopic transverse response. Within the intrinsic Berry-phase picture, the anomalous Hall and anomalous Nernst conductivities are governed by Berry curvature and its energy-weighted distribution over the Brillouin zone[12,13]. Because lattice translations only relate equivalent points in reciprocal space, the relevant symmetry constraints can be analyzed in terms of the magnetic point group. As shown in Fig. 1(a), when a magnetic symmetry operation $\boldsymbol{R}$ relates two momenta $\boldsymbol{k}$ and $\boldsymbol{k}'=\boldsymbol{R}\boldsymbol{k}$ and imposes $\Omega(\mathbf{k}') = -\Omega(\mathbf{k})$, their Berry-curvature contributions cancel upon Brillouin-zone integration, requiring $\sigma / \alpha = 0$. Once $\boldsymbol{R}$ is broken, this symmetry-enforced cancellation constraint is removed, allowing the Brillouin-zone-integrated Berry curvature to acquire a finite value and thereby permitting finite anomalous Hall and anomalous Nernst responses. Consequently, an antiferromagnet may host pronounced Berry-curvature hot spots while exhibiting a vanishing Brillouin-zone-integrated Hall or Nernst response. This situation is particularly relevant to compensated noncoplanar antiferromagnets, where finite local geometric contributions can remain macroscopically silent because of symmetry-imposed cancellation[14]. More broadly, recent symmetry analyses of magnetic geometry and quantum geometry have emphasized that real-space magnetic symmetry can impose direct selection rules on momentum-space geometric responses[15]. The central issue is therefore not simply whether Berry curvature is generated, but whether the magnetic symmetry allows it to remain uncompensated.

Most approaches to controlling anomalous transverse transport have focused on modifying the magnetic degree of freedom itself. Different noncollinear configurations can possess distinct magnetic symmetries and consequently different Hall responses, as demonstrated in Mn-based noncollinear antiferromagnets and magnetic antiperovskites[7,8,11]. Such approaches typically change the spin configuration, Néel-vector orientation, or spin canting together with the symmetry that constrains Berry curvature. A conceptually distinct route is to manipulate the crystallographic part of the magnetic symmetry while keeping the magnetic order fixed. Recent work on CrSb has demonstrated that reconstructing crystal symmetry provides an independent route to modify symmetry-allowed anomalous Hall responses, distinct from merely rotating the Néel vector[16]. In a complementary setting, a recent study of the anomalous in-plane Hall effect in $Fe_3Sn$ explicitly showed that breaking selected mirror and rotational symmetries removes a Berry-curvature cancellation constraint and permits a finite Hall response[17]. These developments suggest that crystal symmetry itself can serve as an independent control dimension for anomalous transverse transport.

Nonmagnetic atoms provide a particularly natural route to such symmetry control. Although they carry no local magnetic moments, their crystallographic positions participate in defining the symmetry of the complete magnetic crystal and can therefore determine whether particular electronic responses are allowed or forbidden. This principle has become central to unconventional magnetism, where the arrangement of nonmagnetic atoms can control momentum-dependent spin splitting in compensated antiferromagnets[18-20]. In FeS, for example, the nonmagnetic S sublattice renders time-reversed antiferromagnetic configurations crystallographically inequivalent and thereby permits a spontaneous Hall response[21]. Importantly, spin splitting and anomalous Hall transport obey distinct symmetry requirements and need not coexist in the same antiferromagnet[5,6]. This naturally raises the question of whether the breaking and restoration of such symmetry operations can be controlled through nonmagnetic atomic occupation, thereby providing a route to switch anomalous transverse transport while the underlying magnetic order remains unchanged.

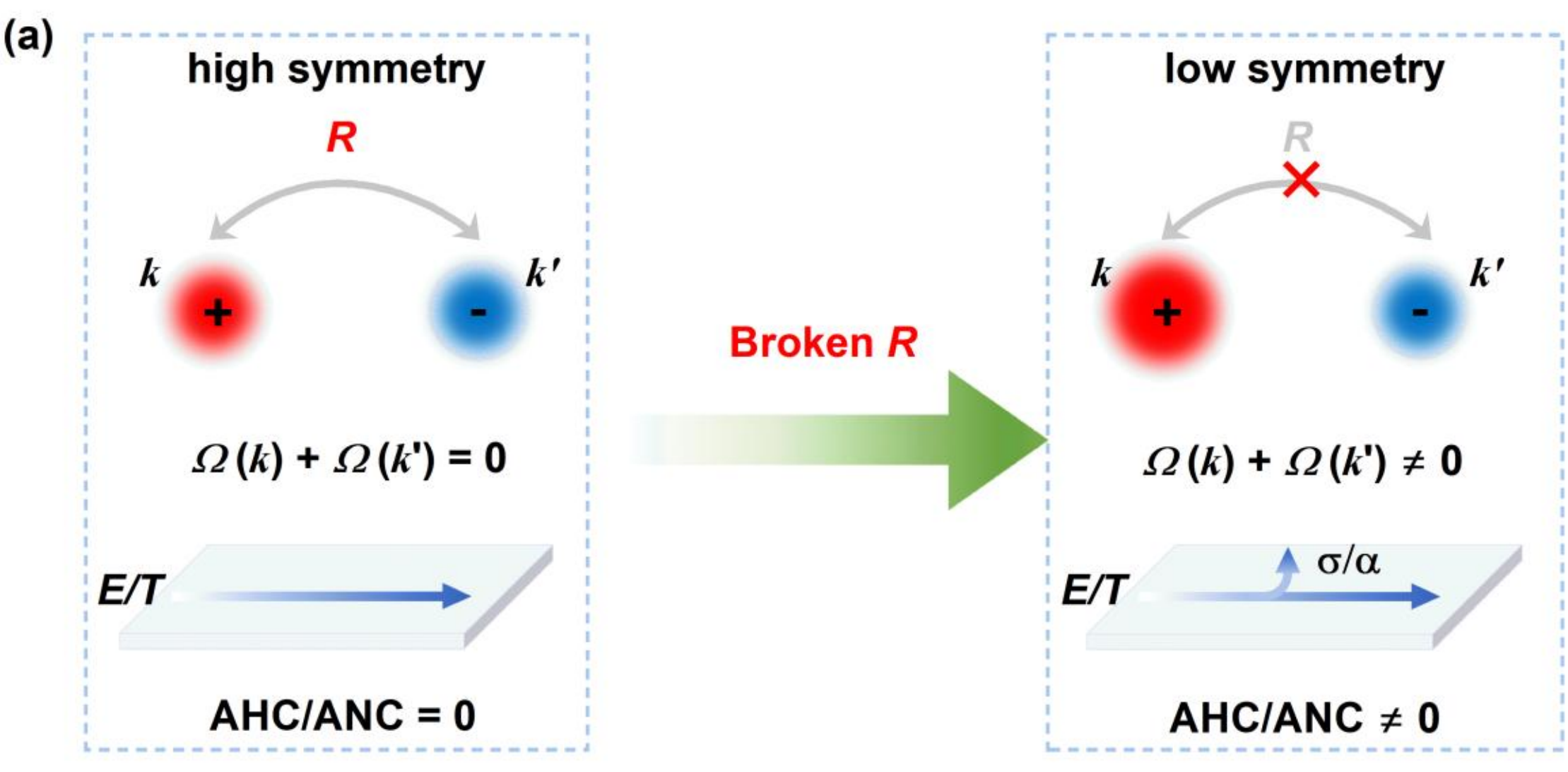

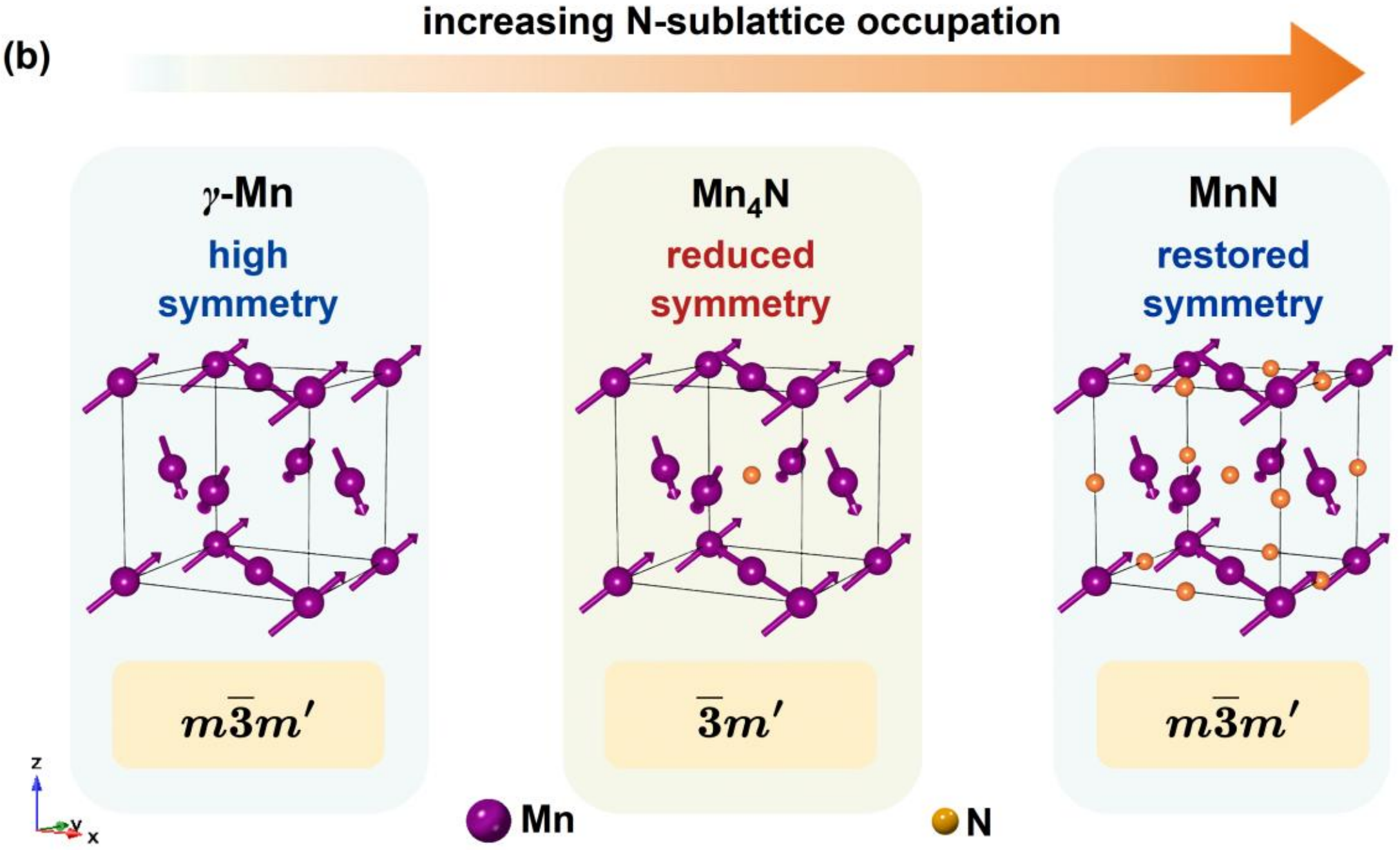

Fig. 1. Symmetry-controlled cancellation of Berry curvature and anomalous transverse transport in compensated antiferromagnets. (a) Schematic illustration of symmetry-enforced Berry-curvature cancellation and its lifting by symmetry breaking. In the high-symmetry state, a symmetry operation $\boldsymbol{R}$ maps $\boldsymbol{k}$ onto a symmetry-related momentum

$\boldsymbol{k}'$ and imposes opposite Berry-curvature contributions, $\Omega(\boldsymbol{k}) + \Omega(\boldsymbol{k}') = 0$, resulting in vanishing Brillouin-zone-integrated Berry curvature and anomalous Hall/Nernst responses. Breaking $\boldsymbol{R}$ removes this cancellation constraint and allows an uncompensated Berry-curvature contribution and finite anomalous transverse transport. The magnetic order remains compensated throughout ($M = 0$). (b) Controlled evolution of the $\gamma$-Mn, $Mn_4N$, and MnN reference states with increasing N-sublattice occupation. The lattice constant and noncoplanar AIAO magnetic configuration are kept identical, while the N-sublattice occupation is varied, driving the magnetic point group from high symmetry to reduced symmetry and back to high symmetry.

Here we address this question using $\gamma$-Mn, $Mn_4N$, and MnN as a controlled series of reference states. The lattice constant and the noncoplanar all-in-all-out Mn magnetic configuration are kept identical, while only the occupation of the nonmagnetic N sublattice is varied. This construction isolates the symmetry effect of nonmagnetic occupation from changes in lattice dimensions or magnetic order. As summarized in Fig. 1(b), N occupation drives a high-symmetry–reduced-symmetry–restored-symmetry sequence of magnetic point groups without changing the Mn magnetic order. Correspondingly, the anomalous Hall and anomalous Nernst responses exhibit the zero–finite–zero evolution. We show that the intermediate $Mn_4N$ state preserves inversion symmetry but loses the relevant twofold rotational symmetries, thereby lifting the symmetry constraint on Berry-curvature cancellation. Our results establish nonmagnetic sublattice occupation as an independent symmetry-control parameter for anomalous transverse transport in compensated antiferromagnets.

## Calculation methods

All first-principles calculations were performed within density functional theory (DFT) using the projector-augmented-wave (PAW) method as implemented in the Vienna *Ab initio* Simulation Package (VASP)[22-24]. The exchange-correlation interaction was described using the Perdew-Burke-Ernzerhof (PBE) functional within the generalized-gradient approximation. A plane-wave cutoff energy of 520 eV was employed, and the electronic self-consistent-field calculations were converged to $10^{-6}$ eV/cell. To establish a controlled comparison among $\gamma$-Mn, Mn(_4)N, and MnN, we constructed a common reference-state framework in which the cubic lattice constant was fixed at a = b = c = 3.85 Å[25] for all three systems. The moments were constrained to the same noncoplanar all-in-all-out (AIAO) antiferromagnetic configuration, while the occupation of the N sublattice was varied from $\gamma$-Mn to the $Mn_4N$ and MnN configurations. Thus, the three systems differ in the occupation of the nonmagnetic sublattice while sharing the same lattice parameter and magnetic configuration. This construction is intended as a controlled theoretical reference rather than as a representation of the respective equilibrium ground states of the three compounds. Spin-orbit coupling (SOC) was included in the self-consistent electronic-structure calculations used for the Berry-curvature and transport analyses. The magnetic point groups of the resulting reference states were determined from the combined crystal and magnetic symmetry of each configuration.

The Bloch states obtained from the DFT calculations were used to construct maximally localized Wannier functions (MLWFs), from which Wannier-interpolated tight-binding Hamiltonians were generated using Wannier90.[26] The projection basis included the *s*, *p*, and *d* orbitals of Mn and the *s* and *p* orbitals of N. The resulting Wannier bands reproduce the corresponding first-principles band structures over the energy range relevant to the transport calculations. The intrinsic AHC ($\sigma^{\gamma}_{\alpha\beta}$, see Formula 2) and ANC ($\alpha^{\gamma}_{\alpha\beta}$, see Formula 3) are calculated from the Berry curvature (Formula 1) using the Wannier-interpolated tight-binding model as proposed by Xiao et al.[13]

$$\Omega_n^\gamma(\boldsymbol{k}) = -\frac{2}{\hbar^2}\operatorname{Im}\sum_{m\neq n}\frac{\left\langle u_n(\boldsymbol{k})\left|\frac{\partial \hat{H}(\boldsymbol{k})}{\partial k_\alpha}\right|u_m(\boldsymbol{k})\right\rangle\left\langle u_n(\boldsymbol{k})\left|\frac{\partial \hat{H}(\boldsymbol{k})}{\partial k_\beta}\right|u_m(\boldsymbol{k})\right\rangle}{\left[E_m(\boldsymbol{k})-E_n(\boldsymbol{k})\right]^2} \quad (1)$$

$$\sigma_{\alpha\beta}^\gamma = -\frac{e^2}{\hbar}\int\frac{d\boldsymbol{k}}{(2\pi)^3}\sum_n f_n(\boldsymbol{k})\Omega_n^\gamma(\boldsymbol{k}) \quad (2)$$

$$\alpha_{\alpha\beta}^\gamma = -\frac{e}{T\hbar}\int\frac{d\boldsymbol{k}}{(2\pi)^3}\sum_n\left\{(E_n-E_f)f_n(\boldsymbol{k})+k_B T\ln(1+e^{\frac{E_n-E_f}{-k_B T}})\right\}\Omega_n^\gamma(\boldsymbol{k}) \quad (3)$$

where $e$ is the elementary charge, $\hbar$ the reduced Planck constant, $k_B$ the Boltzmann constant, $E_f$ the Fermi level, $n$ is the band index, $\alpha,\beta,\gamma = x,y,z$ ($\alpha\neq\beta\neq\gamma$), $E_n$ the band energy, $u_n(\boldsymbol{k})$ is the Bloch function, $f_n(\boldsymbol{k})$ the Fermi-Dirac distribution function, and $\Omega_n^\gamma$ the Berry curvature. We used a $101\times101\times101$ mesh for integrations over BZ to calculate the $\sigma_{\alpha\beta}^\gamma$ and $\alpha_{\alpha\beta}^\gamma$ in WannierTools[27]. Note that this mesh was carefully checked to ensure convergence.

## 3. Results and discussion

To examine this question in a controlled manner, we construct $\gamma$-Mn, $Mn_4N$, and MnN as a common set of reference states, fixing the lattice constant at 3.85 Å and the magnetic configuration to the same noncoplanar all-in-all-out (AIAO) state while varying only the occupation of the N sublattice. This construction is not intended to represent the equilibrium state of each compound, but to eliminate differences associated with lattice relaxation and magnetic reconstruction and thereby provide a controlled comparison of different atomic occupations. As illustrated in Fig. 1(b), the corresponding MPG changes from $m\bar{3}m'$ in γ-Mn to $\bar{3}m'$ in $Mn_4N$ and then returns to $m\bar{3}m'$ in MnN.

Having established that the N-sublattice occupation changes the magnetic symmetry of the reference states, we first examine whether this change is accompanied by momentum-dependent spin splitting. Notably, none of the three reference states exhibits momentum-dependent spin splitting, as confirmed by both self-consistent and non-self-consistent calculations. Thus, varying the N occupation does not simply activate a momentum-dependent spin-splitting response within the fixed AIAO magnetic configuration. We therefore turn to the anomalous transverse responses to determine whether a distinct and symmetry-dependent transport evolution emerges. For this purpose, Wannier-interpolated tight-binding Hamiltonians were constructed from the first-principles electronic structures; the resulting bands closely reproduce the corresponding DFT bands for $\gamma$-Mn, $Mn_4N$, and MnN as shown in Figs. 2(b–d), providing a reliable basis for the subsequent Berry-curvature and transport calculations.

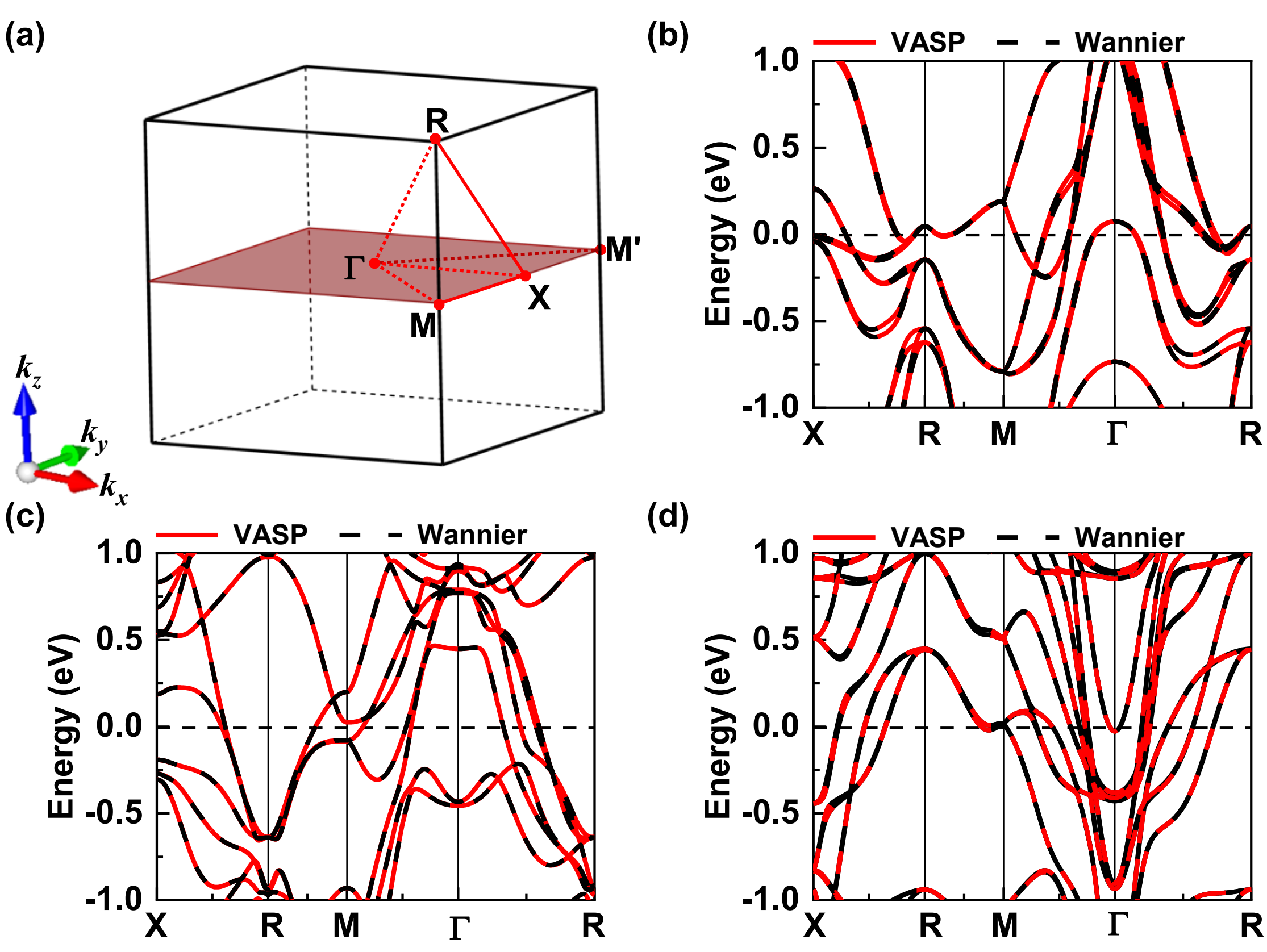

Fig. 2. (a) The first Brillouin zone (black) of the primitive unit cell and its high-symmetry points. The red lines indicate the high-symmetry path plotted in the energy band. The red rectangular plane defines the region used to calculate the Berry curvature shown in Fig. 4. (b)-(d) Energy bands obtained from first-principles calculations (red) and Wannier interpolation (black) for (b) $\gamma$-Mn, (c) $Mn_4N$, and (d) MnN.

A striking contrast emerges in the anomalous transverse responses of the three reference states. Both the anomalous Hall conductivity $\sigma_{(111)}$ and anomalous Nernst conductivity $\alpha_{(111)}$ exhibit a nonmonotonic zero–finite–zero evolution as the N-sublattice occupation increases from γ-Mn to $Mn_4N$ and then to MnN. Near the Fermi level, $\sigma_{(111)}$ is essentially absent in $\gamma$-Mn and MnN, but reaches -126 S/cm in $Mn_4N$, accompanied by a finite $\alpha_{(111)}$ of 0.77 V/mK. Thus, increasing the N occupation does not simply activate anomalous transverse transport; instead, a finite response occurs only for the intermediate $Mn_4N$ reference state. This nonmonotonic behavior indicates that the presence of nonmagnetic atoms alone is insufficient to determine the transverse response, motivating a closer examination of the underlying Berry-curvature evolution.

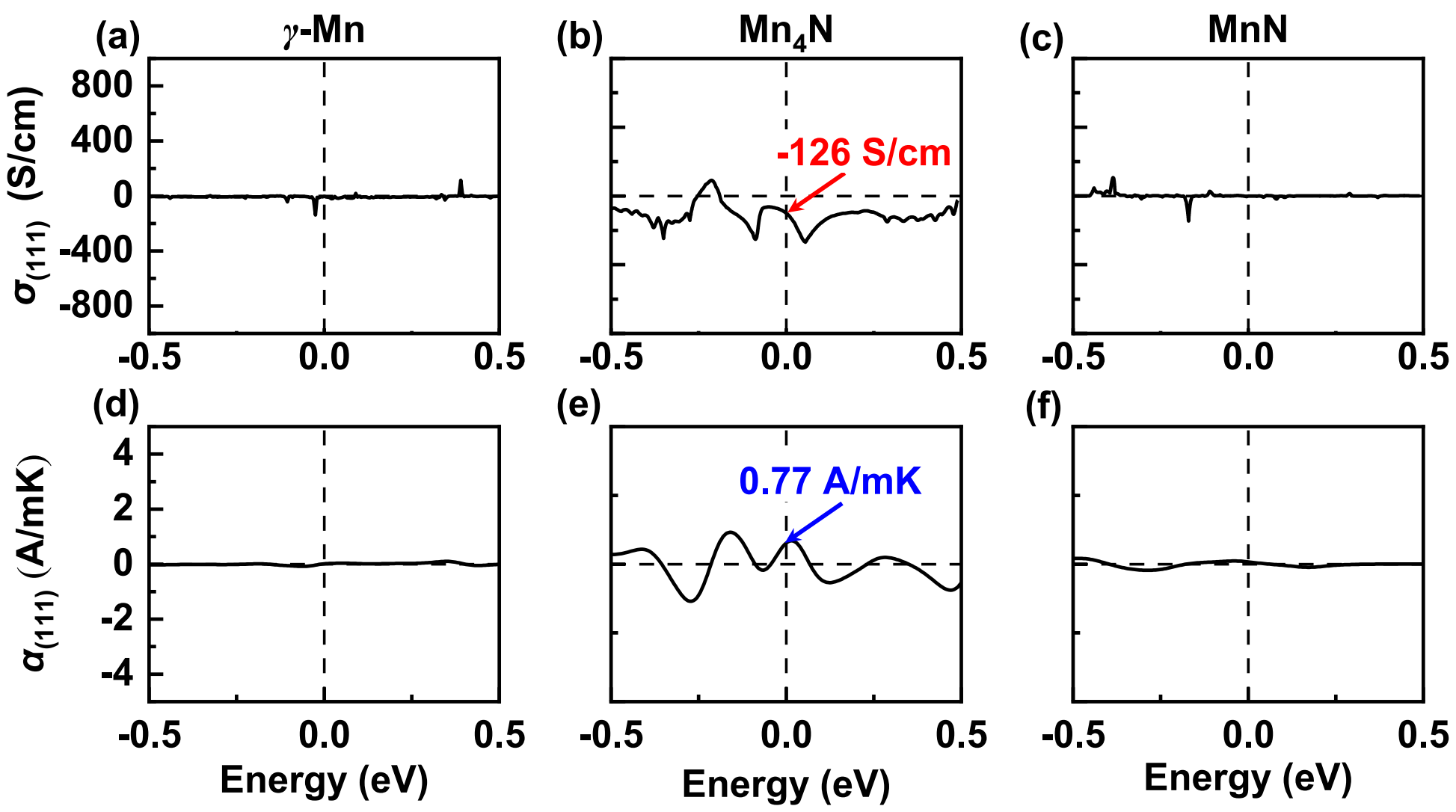

Fig. 3. Calculated anomalous Hall conductivity (AHC) $\sigma_{(111)}$ (a-c) and anomalous Nernst conductivity (ANC) $\sigma_{(111)}$ (d-f) as a function of energy for $\gamma$-Mn (left panels), $Mn_4N$ (middle panels), MnN (right panels).

To trace the microscopic origin of the nonmonotonic transport evolution, we next examine the Berry curvature in momentum space. As shown in Fig. 4(a), the bands of $\gamma$-Mn along the M-Γ-M' path exhibit two nearly symmetric gap openings on opposite sides of Γ. Figure 4(b) further shows that the corresponding Berry curvature displays peaks with opposite signs and equal magnitudes at these two regions. As illustrated in Fig. 4(c), the two-dimensional Berry-curvature distribution in the (001) plane also exhibits a symmetric arrangement of positive and negative contributions, consistent with the opposite-sign Berry curvature along the M-Γ and Γ-M' segments. These contributions therefore cancel upon Brillouin-zone integration, consistent with the vanishing $\sigma_{(111)}$ and $\alpha_{(111)}$ observed in Fig. 3. A similar cancellation behavior is found in MnN. As shown in Figs. 4(g)–4(i), the Berry-curvature distribution again exhibits symmetry-related positive and negative contributions that cancel upon Brillouin-zone integration.

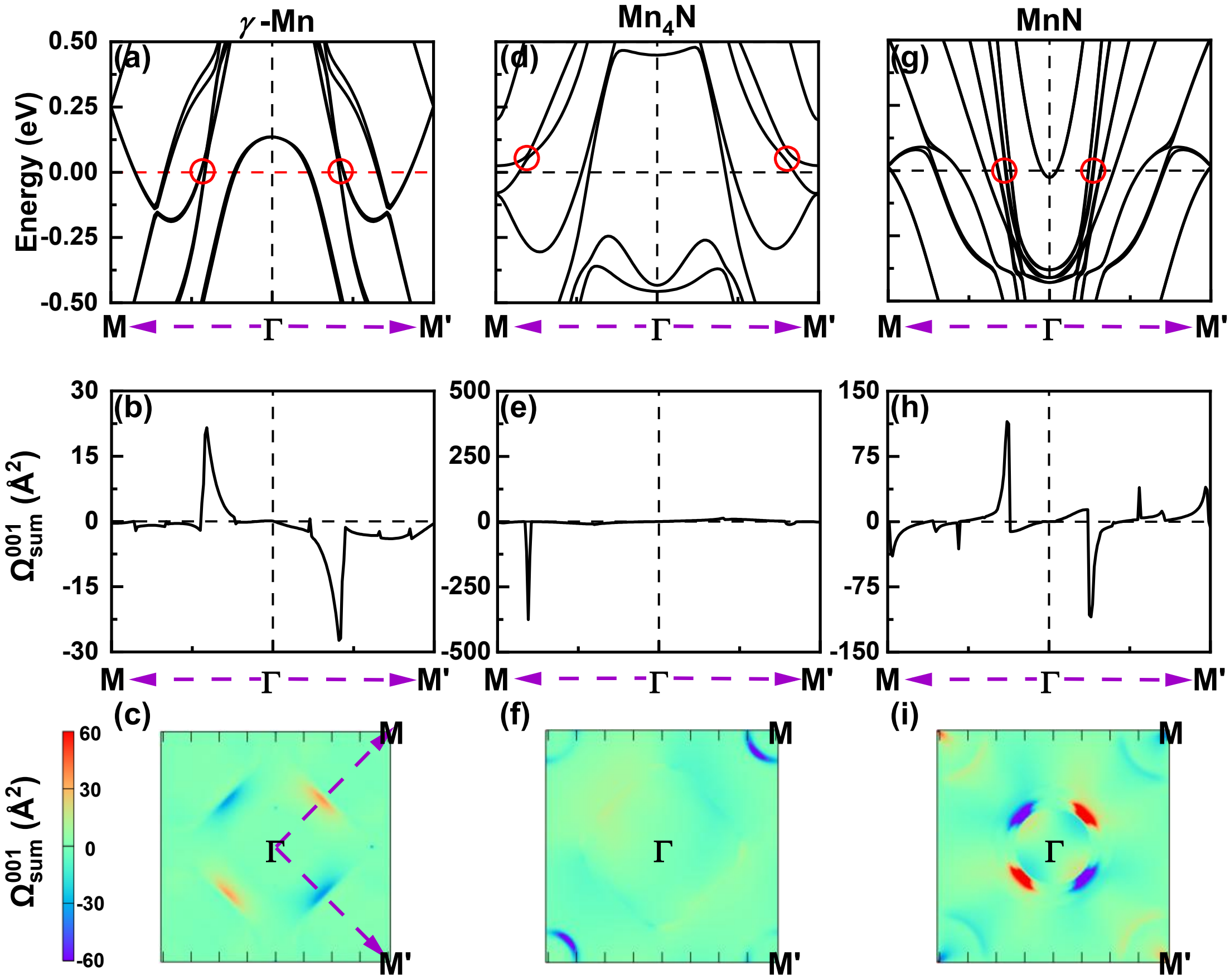

Fig. 4. Energy band of (a) $\gamma$-Mn, (d) $Mn_4N$, and (g) MnN within ±0.5 eV of the Fermi surface. Berry curvature $\Omega_{sum}^{001}$ along the high-symmetry path for (b) $\gamma$-Mn, (e) $Mn_4N$, and (h) MnN, and the color map of $\Omega_{sum}^{001}$ in the (001) for (c) $\gamma$-Mn, (f) $Mn_4N$, and (i) MnN. (c) The path of the M-Γ-M is shown by the purple arrow.

A qualitatively different behavior emerges in $Mn_4N$. As shown in Fig. 4(d), the gap openings on the two sides of Γ along the same M-Γ-M' path become strongly asymmetric, with a narrow gap along the M-Γ segment and a much wider gap along Γ-M'. Figure 4(e) shows that the Berry curvature is dominated by a pronounced negative peak near the narrow-gap region, without a corresponding compensating contribution of comparable magnitude on the opposite side. The two-dimensional Berry-curvature distribution shown in Fig. 4(f) likewise becomes asymmetric, with the finite Berry curvature concentrated around the M-Γ region. Consequently, the positive and negative Berry-curvature contributions no longer cancel completely upon Brillouin-zone integration, leaving a finite net Berry-curvature contribution consistent with the nonzero $\sigma_{(111)}$ and $\alpha_{(111)}$ obtained for $Mn_4N$.

These contrasting Berry-curvature distributions provide the momentum-space origin of the zero–finite–zero transport evolution. The remaining question is which symmetry changes associated with N occupation enforce the cancellation in γ-Mn and MnN and permit its breakdown in $Mn_4N$.

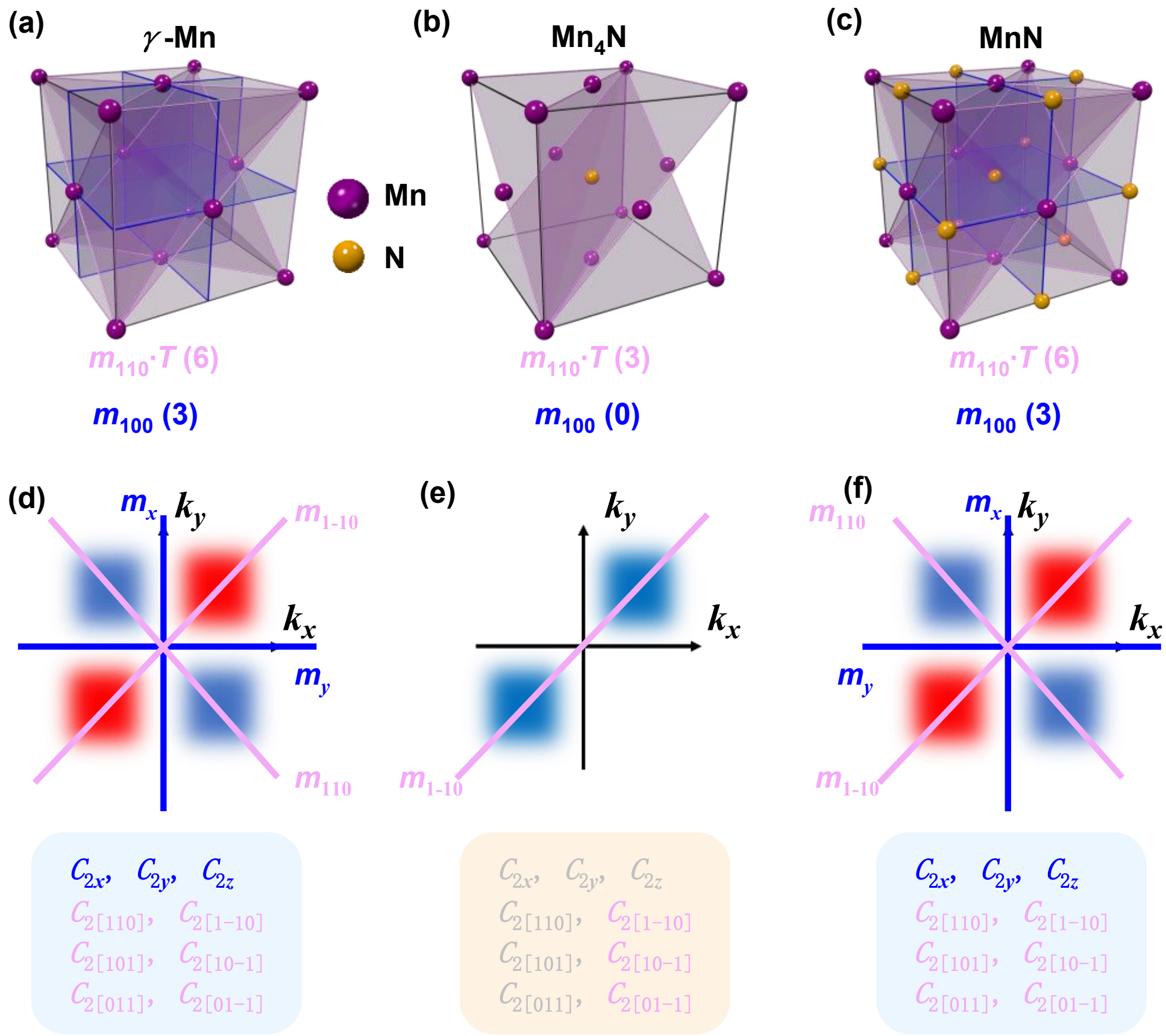

**Fig. 5.** Magnetic-symmetry evolution and the associated Berry-curvature constraints in the γ-Mn, $Mn_4N$, and MnN reference states. (a) The high-symmetry γ-Mn reference state contains three mutually perpendicular crystallographic mirrors $m_{100}$ and six

magnetic mirrors $m_{110} \cdot T$. (b) N occupation reduces the symmetry in $Mn_4N$, removing all three $m_{100}$ mirrors while retaining three $m_{110} \cdot T$ mirrors. (c) Further N occupation restores the high-symmetry in MnN. (d)–(f) Corresponding Berry-curvature distributions in momentum space and their symmetry-related mirror operations for $\gamma$-Mn, $Mn_4N$, and MnN, respectively. The $m_{100}$-type mirrors are the key symmetry operations responsible for the cancellation of the relevant Berry-curvature contributions, whereas the surviving $m_{110} \cdot T$ mirrors in $Mn_4N$ constrain the net Berry curvature to the [111] direction.

Figure 5. Magnetic mirror symmetries in real space and their corresponding Berry-curvature distributions in momentum space for γ-Mn (a, d), $Mn_4N$ (b, e), and MnN (c, f). The blue and pink planes in (a-c) represent the crystallographic mirrors of $m_{100}$ family and the magnetic mirrors of $m_{110} \cdot T$ family, respectively. The numbers in parentheses indicate the number of equivalent mirror planes. The red and blue regions represent positive and negative Berry curvature values, respectively.

Symmetry has been widely recognized as a key factor governing observable electronic and transport responses in magnetic systems. In $Mn_3Ir$, the face-centered-cubic stacking of kagome layers breaks the $m_{111}$ mirror symmetry, enabling a large anomalous Hall conductivity in the presence of spin-orbit coupling.[7] In $Mn_3Sn$, by contrast, the $m_{110}$ mirror symmetries constrain the allowed anomalous Hall conductivity components[8] In altermagnets the crystal rotation symmetry breaks the combined spatial inversion and time reversal *PT* symmetry driving a momentum dependent alternating spin splitting.[18,19] Together, these studies demonstrate that specific crystal and magnetic symmetry operations can determine whether a given electronic or transport response is allowed, motivating us to identify the corresponding symmetry constraints in the present reference-state series.

Having established the contrasting Berry-curvature distributions in Fig. 4, we now identify the symmetry operations that enforce their cancellation or allow a net contribution. Because lattice translations map $\boldsymbol{k}$ onto equivalent points in reciprocal space, the symmetry constraints on the Brillouin-zone-integrated Berry curvature can be analyzed at the level of the MPG. In the present Mn-based reference states, the key symmetry operations corresponding to $\boldsymbol{R}$ are the three mutually perpendicular crystallographic mirror symmetries $m_{100}$, $m_{010}$, and $m_{001}$. These operations relate symmetry-connected momenta in a manner that enforces cancellation of the corresponding Berry-curvature contributions. The $\gamma$-Mn reference state has the high-symmetry $m\bar{3}m'$ magnetic point group, which contains three mutually perpendicular $m_{100}$ crystallographic mirrors and six $m_{110}\cdot T$ magnetic mirrors. The $m_{100}$ mirror can be represented as the combined operation of spatial inversion $P$ and a twofold rotation $C_{2x}$. Under spatial inversion $P$, the momentum transforms as $\boldsymbol{k} \rightarrow -\boldsymbol{k}$, while the Berry curvature, as a pseudovector, remains unchanged. The subsequent $C_{2x}$ rotation reverses the $y$ and $z$ components of the Berry-curvature pseudovector while leaving its $x$ component unchanged. Consequently, states related by $m_{100}$ contribute opposite $y$ and $z$ components of the Berry curvature, whereas the $x$ component retains the same sign. Their integration over the entire Brillouin zone therefore cancels the $y$ and $z$ components exactly. The three mutually perpendicular $m_{100}$, $m_{010}$, and $m_{001}$ mirrors collectively enforce cancellation of all components of the Brillouin-zone-integrated Berry curvature. This symmetry-enforced cancellation accounts for the vanishing $\sigma_{(111)}$ and $\alpha_{(111)}$ obtained for $\gamma$-Mn in Fig. 3.

Introducing interstitial N atoms at the 1b sites selects the [111] body diagonal as the principal trigonal axis and reduces the magnetic point group from $m\bar{3}m'$ to $\bar{3}m'$. This symmetry reduction removes all three standard mirrors and three of the six magnetic mirrors, as indicated by the dashed lines in Fig. 5(b), leaving only the three magnetic mirrors that intersect along the [111] axis. The loss of the $m_{100}$ mirror symmetry does not, however, uniquely specify the underlying symmetry breaking, since $m_{100}$ can be represented as the combined operation of spatial inversion $P$ and a twofold rotation $C_{2x}$. Thus, the loss of $m_{100}$ may result from the breaking of $P$, the breaking of $C_{2x}$, or the breaking of both. Breaking inversion symmetry $P$ removes the inversion relation between $\boldsymbol{k}$ and $-\boldsymbol{k}$ and can allow inversion-asymmetric Berry-curvature distributions and nonlinear Hall responses.[28-31] In contrast, when $P$ is preserved but $C_{2x}$ is broken, the inversion constraint remains while the rotational symmetry associated with $C_{2x}$ is removed. Such rotational-symmetry breaking has been associated with anomalous in-plane Hall effects (IPHEs) in previous studies.[32,33]

The present $Mn_4N$ reference state corresponds to the latter case: the $\bar{3}m'$ magnetic point group preserves inversion symmetry $P$, while the $m_{100}$ mirror and the corresponding $C_{2x}$ rotation symmetry are broken. Unlike the field-induced symmetry breaking reported in $Fe_3Sn$,[17] the symmetry reduction here arises intrinsically from nonmagnetic atomic occupation. Notably, in this work the three $m_{100}$-type mirrors are the key symmetry operations associated with the Berry-curvature cancellation, whereas the surviving $m_{110}{\cdot}T$ mirrors in $Mn_4N$ further constrain the direction of the uncompensated macroscopic Berry curvature. This symmetry-imposed constraint is consistent with the localized and asymmetric Berry-curvature distribution around the M-Γ region shown in Fig. 4. The resulting uncompensated Berry-curvature contribution gives rise to the finite macroscopic anomalous Hall conductivity of -126 S/cm for $Mn_4N$. With further increase in N concentration, the rock-salt MnN phase restores the high-symmetry $m\bar{3}m'$ magnetic point group, reinstating the complete Berry-curvature cancellation and suppressing the anomalous transverse transport responses.

## 4. Conclusions

We have demonstrated that nonmagnetic N-sublattice occupation can control anomalous Hall and Nernst responses without altering the underlying noncoplanar Mn magnetic order. Using controlled $\gamma$-Mn, $Mn_4N$, and MnN reference states with a common lattice constant and identical all-in-all-out magnetic configurations, we isolate the effect of nonmagnetic occupation on magnetic symmetry and Berry-curvature-driven transport. The anomalous transverse responses exhibit a nonmonotonic zero–finite–zero evolution across the series. This behavior originates from an occupation-driven reduction and restoration of magnetic point-group symmetry: the high-symmetry $\gamma$-Mn and MnN states enforce complete cancellation of the Brillouin-zone-integrated Berry curvature, whereas in $Mn_4N$ inversion symmetry remains preserved while the relevant twofold rotational symmetries are broken, lifting the cancellation constraint and allowing a finite net contribution along [111]. Our results establish a direct connection between nonmagnetic sublattice occupation, magnetic-crystal symmetry, Berry-curvature cancellation, and anomalous transverse transport, and identify nonmagnetic occupation as an independent symmetry-control parameter in compensated antiferromagnets.

## Acknowledgements

This work was supported by the National Natural Science Foundation of China (grant no. 51901067, 51971087, 52101233, and 52071279), the Natural Science Foundation of Hebei Province (grant no. E2019205234), the Science and Technology Research Project of Hebei Higher Education (grant no. QN2019154), and the Science Foundation of Hebei Normal University (grant no. L2019B11), and the "333 Talent Project" of Hebei province (grant No. C20231105) and the Science Foundation of Hebei Normal University, China (grant No. L2024B08).